\documentclass[12pt,letterpaper]{article}
\pdfoutput=1

\usepackage{graphicx,array}
\usepackage{url}
\usepackage{color}
\usepackage{latexsym}
\usepackage{amsthm}
\usepackage{amsmath}
\usepackage{amssymb}
\usepackage{amsfonts}
\usepackage[numbers,sort&compress]{natbib}
\usepackage{bm}
\usepackage{slashed}
\usepackage{mathrsfs}
\usepackage{enumerate}
\usepackage{tikz}
\usepackage{siunitx}
\usepackage{mdframed}
\usepackage{lipsum}
\usepackage{setspace}

\usepackage[version=4]{mhchem}

\usepackage[utf8x]{inputenc}%
\usepackage{tcolorbox}%

\usepackage{hyperref} %Automatically links \label and \ref commands; Always load last
\hypersetup{
    colorlinks=true,       % false: boxed links; true: colored links
    linkcolor=red,          % color of internal links
    citecolor=blue,        % color of links to bibliography
    filecolor=magenta,      % color of file links
    urlcolor=blue           % color of external links
}
\usepackage[all]{hypcap} %Link navagates to top of figure instead of caption (below fig)

\usepackage{natbib}
\newcommand{\nc}{\newcommand}

\nc{\beq}{\begin{equation}}  
\nc{\eeq}{\end{equation}}  
\nc{\beqa}{\begin{eqnarray}}  
\nc{\eeqa}{\end{eqnarray}}  

\def\be{\begin{equation}}
\def\ee{\end{equation}}
\def\bea{\begin{eqnarray}}
\def\eea{\end{eqnarray}}

\nc{\bit}{\begin{itemize}}  
\nc{\eit}{\end{itemize}}

\def\sol/{\textrm{DM soliton}}
\def\sols/{\textrm{DM solitons}}
\def\Tittlesol/{\textrm{DM Soliton}}
\def\Tittlesols/{\textrm{DM Solitons}}
\def\EDB/{\textrm{EWS-DMB}}
\def\EDBs/{\textrm{EWS-DMBs}}
\def\DMB/{\textrm{DMB}}
\def\DMBs/{\textrm{DMBs}}
\def\DMN/{\textrm{DM number}}
\def\DMNs/{\textrm{DM numbers}}

\DeclareRobustCommand\encircle[1]{%
  \tikz[baseline=(X.base)] 
    \node (X) [draw, shape=circle, inner sep=-1] {\strut \raisebox{-0.5pt}[0pt]{#1}};}
\newcommand{\Pcirc}{\encircle{$\Phi$}}
\newcommand{\Msol}{M_{\tiny \Pcirc}}
\newcommand{\Rsol}{R_{\tiny \Pcirc}}
\newcommand{\rhosol}{\rho_{\tiny \Pcirc}}
\newcommand{\PNcirc}{\encircle{{\footnotesize{$\Phi$}}{\tiny{\text{N}}}}}
\usepackage{floatrow}
\newfloatcommand{capbtabbox}{table}[][\FBwidth]

\usepackage{blindtext}

\title{ \bf
\raggedleft {\small\normalfont PITT-PACC-1912}\\\vspace{0.5cm}
Nucleus Capture by Macroscopic Dark Matter
\author{\large Yang Bai$^{\,\star}$ and Joshua Berger$\,^\diamond$}
\date{\small  \it 
$^\star$Department of Physics, University of Wisconsin-Madison, Madison, WI 53706, USA \\
$^\diamond$Department of Physics and Astronomy, University of Pittsburgh, Pittsburgh, PA 15260, USA  \\
}
}

\begin{document}

\maketitle

\setlength{\parskip}{0.2ex}

\begin{abstract}
For a class of macroscopic dark matter with a large interaction strength with 
Standard Model particles, a nucleus could be captured by the dense, heavy dark 
matter as it traverses ordinary material. The radiated photon carries most of 
the binding energy and is a characteristic signature for dark matter detection. 
We develop analytic formulas and present numerical results for this radiative 
capture process  in the low energy, non-dipole limit. Large-volume neutrino 
detectors like NO$\nu$A, JUNO, DUNE and Super(Hyper)-K may detect multi-hit or 
single-hit radiative capture events and can search for dark matter up to one 
gram in mass.
\end{abstract}

\thispagestyle{empty}  
\newpage  
  
\setcounter{page}{1}  

\newpage

%\hypersetup{linktocpage} % the table of content can have hyperlink
%\tableofcontents 
%\hypersetup{linkcolor=red} %after fix black color for table of content and change the link color

%\newpage

%%%%%%%%%%%%%%%%%%%%%%%%%%%%%%%%%%%%%
% Introduction
%%%%%%%%%%%%%%%%%%%%%%%%%%%%%%%%%%%%%
\section{Introduction} 
\label{sec:intro}
%%%%%%%%%%%%%%%%%%%%%%%%%%%%%%%%%%%%%
Macroscopic dark matter (MDM) is a general class of  models with dark matter (DM) in 
a compact and composite state with a large radius and mass. The composite 
consists of many elementary dark matter particles and has dramatically different properties from 
a microscopic particle. In the literature, MDM appears in many scenarios either 
within the Standard Model (SM) or beyond. For instance, the massive 
astrophysical compact halo object (MACHO) such as brown dwarfs is made of SM 
particles and has long been proposed as a dark matter 
candidate~\cite{1986ApJ...304....1P,1991ApJ...366..412G}, although their 
abundance is constrained by microlensing experiments to occupy only a small 
fraction of dark matter~\cite{Tisserand:2006zx}. For lighter MDM below the 
microlensing threshold mass around $10^{-11}~M_\odot$~\cite{Niikura:2017zjd}, 
one could have so-called quark nuggets that are made of quark matter in the 
unconfined QCD phase~\cite{Witten:1984rs,Liang:2016tqc,Bai:2018vik}. These 
objects have a QCD scale energy density, $\sim 10^{15}\,\mbox{g}/\mbox{cm}^3$, 
and a radius $\sim0.01~{\rm cm}$. Beyond the SM, MDM can be composed of bosonic 
constituents, as in non-topological 
solitons~\cite{Friedberg:1976me,Coleman:1985ki,Kusenko:1997si}. For example, in 
Ref.~\cite{Ponton:2019hux}, electroweak symmetric dark matter balls (\EDB/) 
were proposed as a MDM candidate in the simple Higgs-portal dark matter model 
with an electroweak scale energy density, $\sim 10^{27}\,\mbox{g}/\mbox{cm}^3$ 
and a radius $\sim 10^{-8}$~cm. Other than quark nuggets or \EDBs/ with 
well-defined QCD or electroweak interactions with SM particles, there are also 
other MDM models with only gravitational interactions or other unknown 
interactions with ordinary matter, including dark quark 
nuggets~\cite{Bai:2018dxf}, asymmetric dark matter 
nuggets~\cite{Gresham:2017zqi} and dark blobs~\cite{Grabowska:2018lnd} (for a 
recent review see~\cite{Jacobs:2014yca}). 

Some MDM candidates like QCD quark nuggets and \EDBs/ have relatively large 
interaction cross sections with ordinary matter. QCD quark nuggets behave 
basically like a very heavy nucleus. When they elastically scatter off a 
nucleus on the target, various nearby bound states of MDM and nucleus enhance 
the scattering process, leading to a geometric size cross section at large 
radius. Similarly, for \EDBs/, the nucleus  has slightly different masses 
inside and outside the dark matter state. For a large radius $\Rsol$ above the 
bound state threshold radius, the elastic scattering cross section saturates 
the geometric one and varies between $2 \pi \Rsol^2$ and $4 \pi \Rsol^2$ up to 
quantum mechanical shadowing effects~\cite{Ponton:2019hux}. This elastic 
scattering cross section is rather large such that the MDM may interact with 
nuclei multiple times in a detector.  Traditional dark matter direct detection 
experiments that look for a single-hit event may veto the MDM-induced multi-hit events 
and may not be suitable to search for ``strongly interacting'' 
MDM~\cite{Bramante:2018qbc,Bramante:2018tos}. Because of the spectacular multi-hit signature, potential experiments searching for MDM do not necessarily need 
to be located underground. The key requirement is to have a large volume 
detector to compensate the smallness of the MDM flux. Indeed, the large 
neutrino detectors like Borexino, ICARUS, NO$\nu$A, JUNO,  Super-Kamiokande (Super-K), DUNE, Hyper-Kamiokande (Hyper-K) or even IceCube 
could be used to search for MDM. On the hand, the trigger thresholds of the 
neutrino detectors become crucial because the individual MDM scatterings do not 
deposit that much energy. For instance, Borexino and JUNO could have a 
sufficiently low threshold energy to detect some multi-hit scattering 
events~\cite{Ponton:2019hux}, but not for the other larger volume 
experiments, which have energy thresholds of at least 1 MeV. 

In this paper, rather than studying elastic scattering events of MDM, we point 
out another interesting MDM-induced signature. As MDM hits a nucleus in the 
detector, the nucleus and MDM could form one of many bound states, as mentioned 
in Ref.~\cite{Ponton:2019hux} for \EDBs/. Just like hydrogen formation from an 
electron and a proton, $e^-  + p \rightarrow \mbox{H} + \gamma$, the radiative 
capture or recombination process can generate a photon in the final state. The 
photon carries most of the binding energy and could have an energy higher than 
the kinetic energy of the scattering system. Similarly, the nucleus 
radiative capture process by a MDM, $\ce{^A_Z N} + \mbox{MDM} \rightarrow 
\mbox{bound states} + \gamma$, can produce energetic photons in the final 
state. Depending on the detailed properties of the MDM, the photons produced 
could have an energy above 1 MeV or even up to GeV, which could be observed by 
a larger volume neutrino detector.

The calculation for the radiative capture cross section can be found in several 
systems. Other than the hydrogen recombination  
process~\cite{Kramers,1978JETP...48..639K} with a long-range Coulomb force, the 
capture of a nucleon or nucleus by a grand unified theory magnetic monopole has 
been studied in~\cite{Olaussen:1983bm}. For both cases, because of the radiated 
photon wavelength is much longer than the Bohr radius, the dipole approximation 
has been used to simplify the calculation. The situation is different for a MDM 
with a hard-sphere structure. The released photon energy $q$ could be so 
energetic such that its wavelength $2\pi/q$ could be much shorter than the 
MDM size $\Rsol$. Therefore, a {\it non-dipole} calculation is needed to study 
the radiative capture process of MDM with a large $\Rsol$. The situation is 
similar to the neutron capture by a nucleus, although the underlying 
interactions are different and one usually relies on numerical tools to 
estimate the cross sections~\cite{neutron-book}. In this paper, without relying 
on dipole approximation, we take 
the {\it low energy} limit with a small scattering momentum $k$ such that $k 
\, \Rsol \ll 2 \pi$, perform an analytic calculation, and extrapolate the 
radiative capture cross section to a large radius beyond this approximation. 

Our paper is organized as follows. We first use the \EDB/ as a concrete MDM 
example to set up the stage for our calculation in Section~\ref{sec:models}.  
Section~\ref{sec:cross-section} contains the main calculations with the dipole 
limit in Subsection~\ref{section:dipole}, low energy limit in 
Subsection~\ref{sec:low-energy} and extrapolation to a large radius in 
Subsection~\ref{sec:large-radius}. The prospects of detection in various 
neutrino detectors are discussed in Section~\ref{sec:detection}. We summarize 
our results in Section~\ref{sec:conclusion}.

%%%%%%%%%%%%%%%%%%%%%%%%%%%%%%%%%%%%%
% Models
%%%%%%%%%%%%%%%%%%%%%%%%%%%%%%%%%%%%%
\section{Example macroscopic dark matter: \EDB/} 
\label{sec:models}
%%%%%%%%%%%%%%%%%%%%%%%%%%%%%%%%%%%%%
We use \EDB/ as a working example to discuss the radiative capture of nuclei by 
macroscopic dark matter. Our analysis can be also applied to other types of 
MDM that have a large interaction strength with SM particles. As discussed in 
Ref.~\cite{Ponton:2019hux}, there exists a non-topological soliton state for 
dark matter in the simple 
Higgs-portal dark matter model with an unbroken $U(1)_\Phi$ dark matter number 
symmetry. The relevant interactions beyond the SM have
\beqa
\mathcal{L} \supset - \lambda_{\phi h} \Phi^\dagger \Phi H^\dagger H - m^2_{\phi,0} \Phi^\dagger \Phi - \lambda_\phi (\Phi^\dagger \Phi)^2 ~,
\eeqa
with $\lambda_{\phi h}$ as the Higgs-portal interaction, $m_{\phi,0}$ as the 
dark matter bare mass independent of electroweak (EW) symmetry breaking, 
$\lambda_\phi$ as the self-interaction of dark matter particle $\Phi$. After 
EW symmetry breaking, the free dark matter particle mass is $m_\phi = 
(m^2_{\phi,0} + \lambda_{\phi h} v_{\rm EW}^2/2)^{1/2}$ with $v_{\rm 
EW}=246$~GeV, the EW vacuum expectation value (VEV). 

In addition to the free particle dark matter state, there is a non-topological 
soliton state of dark matter with a charge $Q = i \int dx^3 ( \Phi^\dagger 
\partial_t \Phi - \Phi \partial_t \Phi^\dagger)$. Solving the classical 
equations of motion for both $\Phi$ and $H$ and in the large $Q$ limit, the 
dark matter soliton state has a mass of
\beqa
\Msol = Q\, \omega_c = Q\, \left[m^2_{\phi, 0} + \left(\lambda_\phi/4 \lambda_h\right)^{1/2} m_h^2 \right]^{1/2} ~.
\eeqa
Here, $\lambda_h\approx 0.13$ is the Higgs quartic coupling in the SM and $m_h 
\approx 125$~GeV is the Higgs boson mass. In the parameter region with 
$\sqrt{\lambda_\phi}/\lambda_{\phi h} < 1.4$, one has $\omega_c < m_\phi$, so 
the soliton state has a lighter mass per charge than a free dark matter particle state. 
For a non-negligible $\lambda_\phi$ and a spherically symmetric \EDB/, the 
self-interaction of $\Phi$ field induces a step-like or hard-sphere profile for 
the $\Phi$ field up to a radius $\Rsol$ and a wall thickness of $1/v_{\rm EW}$. 
In the 
large $Q$ limit, there are simple scaling laws between the \DMBs/ charge, size 
and mass: $\Msol \sim Q \sim \Rsol^3$. The energy density of a \DMB/ is 
\beqa
\rhosol = \frac{\Msol}{(4\pi/3) \Rsol^3} \sim v_{\rm EW}^4 \sim  \left( 100 ~\mbox{GeV}\right)^4 ~,
\eeqa
which is much denser than ordinary matter. The early universe production of the 
\EDB/ from a first-order phase transition has also been discussed in 
Ref.~\cite{Ponton:2019hux}.  The \EDB/s can have a macroscopic mass above 1 
gram and a radius above $10^5\,\mbox{GeV}^{-1}$, dramatically above the 
electroweak scale. The masses and radii of \EDB/s are very sensitive to the 
portal coupling strength $\lambda_{\phi h}$, which is also responsible for 
providing the first-order electroweak phase transition. In the range of 
$\lambda_{\phi h}$ from 2 to 9, the average DMB mass ranges from $1\times 
10^{24}$~GeV and $9\times 10^{33}$~GeV~\cite{Ponton:2019hux}. There is no 
collider constraint on the model parameter space with $\lambda_{\phi h} > 2$. 
For a large coupling, the $\Phi$ particle has a mass above half the Higgs mass 
and has a suppressed pair-production cross section from an off-shell Higgs 
boson at the Large Hadron Collider.

Due to the interplay of $\Phi$ and Higgs profiles, the field value of $\Phi$ in 
the inner region of \EDB/ is large enough to flip the sign of the effective Higgs mass 
squared, $\lambda_{\phi h} \Phi^\dagger \Phi - \lambda_h v_{\rm EW}^2$, and 
prefers a zero Higgs VEV or unbroken electroweak symmetry. Hence, this soliton 
state is an interesting macroscopic dark matter, because it sustains an EW 
symmetric ``vacuum'' in a finite region of space, immersed in the normal EW 
breaking vacuum. 

When \DMB/ with a large radius scatters with a nucleon or a nucleus, a large 
scattering cross section is generically anticipated. For elastic scattering, 
there are effects due to shallow bound states at several partial waves. After 
summing over these partial waves, the cross section follows a ``hard ball" 
behavior, between 2 and 4 times the geometric cross 
section~\cite{Ponton:2019hux}. Multi-hit signals are the characteristic 
features of the \DMB/ elastic scattering events. Since only $\mathcal{O}(10~\mbox{keV})$ are anticipated from each scattering, a low energy threshold below around 1 MeV is required to identify the dark matter scattering events. In this paper, we will instead concentrate on the important radiative capture process, which can 
convert the binding energy of a nucleus and a \DMB/ into photons with energies 
of $\mathcal{O}(1~\mbox{MeV}-100~\mbox{MeV})$, depending on nucleus mass 
number. 

%%%%%%%%%%%%%%%%%%%%%%%%%%%%%%%%%%%%%
% Scattering
%%%%%%%%%%%%%%%%%%%%%%%%%%%%%%%%%%%%%
\section{Radiative capture cross section}
\label{sec:cross-section}
%%%%%%%%%%%%%%%%%%%%%%%%%%%%%%%%%%%%%
A nucleus can be captured by a \DMB/ while emitting a photon in a process 
similar to neutron radiative capture by a nucleus, such as $n + 
\ce{^{197}_79Au} \rightarrow \ce{^{198}_79Au}+ \gamma$. Explicitly, the 
\DMB/-induced radiative capture process is 
\beqa
\ce{^A_Z N} + \Pcirc \rightarrow  \PNcirc +  \gamma ~,
\eeqa
with $\PNcirc$ as a bound state of \DMB/ and a nucleus. For the neutron capture 
case, depending on the incident neutron kinetic energy, there is a low energy 
region with $1/v$ scaling, an intermediate resonant region and a fast neutron 
region. For the \DMB/ case, the relative speed between the \DMB/ and the target 
nucleus in a laboratory is roughly the averaged dark matter speed in our local 
galaxy or around $v \sim 300~\mbox{km}/\mbox{s} \approx 10^{-3}\,c$. Using the reduced mass $\mu \approx A\,m_p$ and the proton mass $m_p = 0.938$~GeV, the kinetic energy of the scattering process is $E_{\rm kin} \approx k^2/(2 \mu) \approx m_A\,v^2/2 \approx A\times 0.5~\mbox{keV}$, which does not vary too much for comparable nucleus mass number. On the other hand, we do not know the exact radius of \DMB/ and will keep the radius as a free parameter of the model. As with neutron radiative capture, we will see three qualitatively different regions: no bound 
state formation, resonant scattering and geometric cross section saturation. 

Though the radiated photon of course requires a relativistic description, both 
the initial scattering state and final bound state are well described by 
non-relativistic mechanics. For a heavy \DMB/ with $\Msol \gg m_A$, the 
center-of-mass frame is approximately the rest frame of \DMB/. Effectively, one 
can think that a nucleus enters the inner region of \DMB/. 

For simplicity, we only consider spherically symmetric \DMBs/ in this work.  
The change in  the profile of the Higgs field from non-zero outside to zero 
inside is rapid, forming a ``wall'' of thickness roughly $\pi/v_{\rm EW}^{-1} \ll 
\Rsol$~\cite{Ponton:2019hux}. This change in the Higgs VEV modifies the 
contribution of the quark masses to the mass of the nucleus, which is to 
excellent approximation the reduced mass of the system. The shift of the 
nucleus mass from outside to inside of the  \DMBs/ is given by
\begin{equation}
\Delta m_A \, \approx \, - A \, y_{hNN} \, v_{\rm EW} ~,
\end{equation}
where $A$ is the mass number of the nucleus and $y_{hNN}$ is the nucleon 
Yukawa coupling, which is roughly the same for protons and neutrons and has a 
value of $1.1 \times 10^{-3}$~\cite{Cheng:2012qr,Alarcon:2012nr}. The shift in 
the mass is small 
since the contribution of the Higgs-generated quark masses to the nucleon 
masses is small compared to that of QCD dynamics. A non-relativistic expansion 
of the Klein-Gordon equation for the nucleus in the background of the changing 
Higgs VEV profile through the spherically symmetric \DMB/ can be performed. 
Note that physical fields of any Lorentz representation 
nevertheless satisfy the 
Klein-Gordon equation. The spin degrees of freedom decouple in the 
non-relativistic limit. Thus, the Schr\"odinger analysis applies to nuclei of 
any spin.

If the expansion is done 
around the outside mass $Z \, m_p + (A-Z)\, m_n$, the changing Higgs VEV 
generates a spherical potential well for the nucleus inside the \DMB/ in the 
``thin wall'' approximation. The potential has a depth of
\begin{equation}
V_0 \, = \, - \Delta m_A ~.
\end{equation}
We allow for some uncertainty in the determination of this effective potential 
due to 
nuclear effects as well as other models different from the baseline  \EDB/ 
model and with a suppressed modification on the nucleus mass. More concretely 
and  in our later numeric calculation, we will use $V_0 \, =\,  A \times 
32~{\rm MeV}$, which is around one eighth of the maximal electroweak restored 
case.
The scattering cross section becomes 
a ``textbook'' one, although we are not aware of the relevant results in the 
large radius parameter region, where the usual dipole approximation breaks 
down. 
In the \DMB/-nucleus center-of-mass frame, the relevant Schr\"odinger equation 
is
\beqa\label{eq:schrodinger}
- \frac{1}{2 \, \mu} \, \nabla^2 \psi(\mathbf{x}) + V(\mathbf{x}) \, \psi(\mathbf{x}) \, = \, E \, \psi(\mathbf{x}), \qquad \quad V(\mathbf{x}) \, = \, \left\{\begin{array}{l l} -V_0 \,, & r \leq \Rsol \, \\ 0 \,, & r > \Rsol\end{array} \right.,
\eeqa
where $r = |\mathbf{x}|$ is the magnitude of the center of mass coordinate. 
%For different nuclei, because of the 
%coherent Higgs scalar coupling to nucleons (ignoring the small nuclear form 
%factor effects), $V_0 \approx ( A \, y_{h N N} \,v_{\rm EW})^2/(2 m_A) \approx 
%A (  y_{h N N} \,v_{\rm EW})^2/(2 m_p) \approx A\times 32$~MeV. Here, the 
%%%coupling of Higgs boson to a 
%nucleon $y_{h N N} \approx 1.1 \times 
%10^{-3}$~\cite{Cheng:2012qr,Alarcon:2012nr}. 

%%%%%%%%%%%%%%%%%%%%%%%%%%%%%%%%%%%%%
% Bound States
%%%%%%%%%%%%%%%%%%%%%%%%%%%%%%%%%%%%%
\subsection{Bound states}
\label{sec:Bound-State}
%%%%%%%%%%%%%%%%%%%%%%%%%%%%%%%%%%%%%
For bound states, i.e.\ states with $E < 0$, we write the wave functions in the 
spherical coordinate with quantum numbers $n, \ell, m$
\beqa
\psi_{n\ell m}(\mathbf{x}) \, = \, R_{n\ell}(r) \, Y_{\ell m}(\hat{\mathbf{x}}) ~.
\eeqa
Here, $Y_{\ell m}(\hat{\mathbf{x}})$ is a spherical harmonic. The bound state 
is normalized as $\int d^3x \, |\psi_{n\ell m}(\mathbf{x})|^2 = 
1$. The normalizable radial wave functions can be expressed in terms of 
spherical Bessel functions
\beqa
R_{n\ell}(r) = \left\{   \begin{array}{l l}  R_{n\ell}^{\rm in}(r) \, = \, d_1 \, j_\ell(\kappa_{n\ell} \, r) \,,  & \quad r \leq \Rsol \,, \\ [1.0em]
    R_{n\ell}^{\rm out}(r) \, = \, d_2 \, [j_\ell(i \, k_{n\ell} \, r) + i\, y_\ell(i \, k_{n\ell} \, r)] \,, &\quad   r > \Rsol  \,,
   \end{array}
\right.
\eeqa
with $\kappa_{n\ell} = \sqrt{2 \, \mu \, (V_0 - |E_{n\ell}|)}$ and $k_{n\ell} = 
\sqrt{2 \, \mu \, |E_{n\ell}|}$. The coefficients $d_{1,2}$ and the energy 
eigenvalues $E_{n\ell}$ are determined by the boundary conditions 
$R_{n\ell}^{\rm in}(\Rsol) \, = \, R_{n\ell}^{\rm out}(\Rsol)$, 
$R_{n\ell}^{\prime\,{\rm in}}(\Rsol) \, = \, R_{n\ell}^{\prime\,{\rm 
out}}(\Rsol)$ and the normalization condition $\int_0^\infty dr \, r^2 \, 
R_{n\ell}^2(r)=1$. While the energy eigenvalue equation cannot be solved 
analytically without any approximations, the coefficients $d_1$ and $d_2$ are 
found to be
\beqa
d_1 \, = \, \frac{1}{N_{n\ell} \, j_\ell(\kappa_{n\ell} \, \Rsol)} ~,\qquad d_2 
\, = \, \frac{1}{N_{n\ell} \, [j_\ell(i \, k_{n\ell} \, \Rsol) + i\, y_\ell(i 
\, k_{n\ell} \, \Rsol)]} ~,
\eeqa
where
\beqa
\label{eq:bound-norm}
N_{n\ell}^2 \, = \, \frac{1}{2} \, \Rsol^3 \, 
\left[\frac{K_{\ell-1/2}(k_{n\ell} \, \Rsol) \, K_{\ell+3/2}(k_{n\ell} 
\Rsol)}{K_{\ell+1/2}^2(k_{n\ell} \Rsol)} - \frac{J_{\ell-1/2}(\kappa_{n\ell} \, 
\Rsol) \, J_{\ell+3/2}(\kappa_{n\ell} \Rsol)}{J_{\ell+1/2}^2(\kappa_{n\ell} 
\Rsol)}\right],
\eeqa
in terms of Bessel functions $J_\nu$ and $K_\nu$.

\begin{figure}[tb!]
\centering
\includegraphics[width=0.475\textwidth]{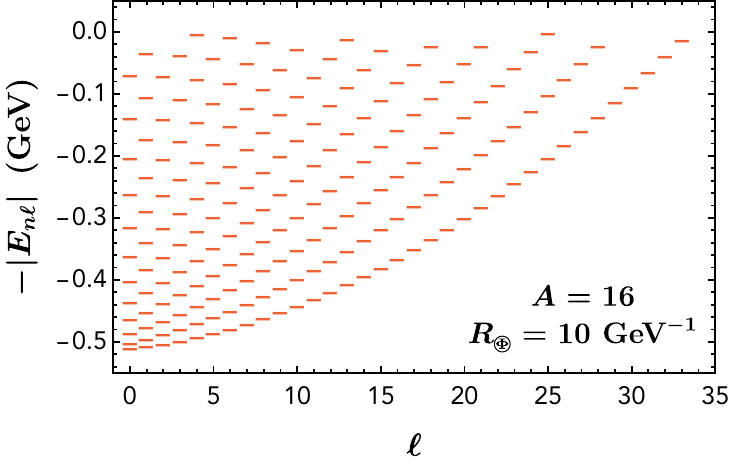}
\hspace{3mm}
\includegraphics[width=0.475\textwidth]{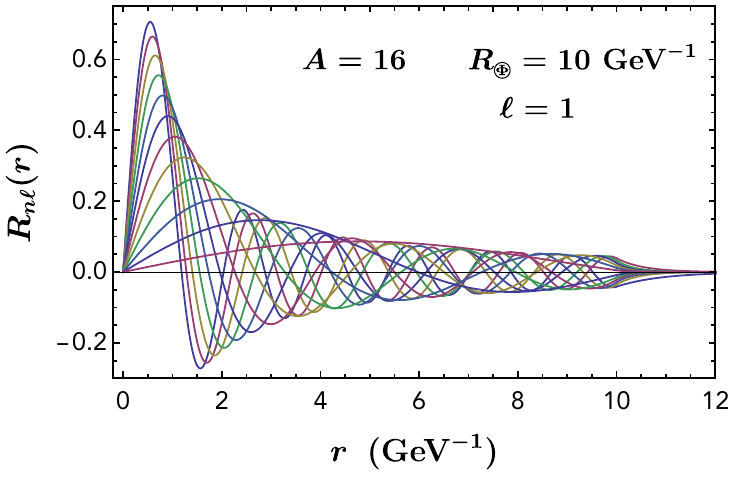}
\caption{Left panel: the energy levels for nucleus-\DMB/ bound states for 
different partial wave numbers $\ell$. Right panel: the 12 radial wave functions 
as a function for $r$ for the $p$-wave bound states with $\ell=1$.}
\label{fig:bound-states}
\end{figure}

For each partial wave $\ell$, there is a threshold radius $R^{\ell}_{\rm th}$ 
below which there are no bound states. The threshold is given by
\beqa
R^{\ell}_{\rm th} = \dfrac{\pi}{2\,\sqrt{2\,\mu\,V_0}} \,J_{\ell -1/2, 1}  \,,
\eeqa
with $J_{\nu,1}$ as the first zero of the Bessel function $J_\nu$. For example, 
one has $R_{\rm th}^0 = 0.41~\mbox{GeV}^{-1}$ and $R_{\rm th}^1 = 
0.80~\mbox{GeV}^{-1}$ for $A=16$. For a large radius $\Rsol$, many bound states exist. For 
$A=16$ and $\Rsol=10~\mbox{GeV}^{-1}$, we show the energy levels in the left 
panel and the radial wave functions for $\ell=1$ in the right panel of 
Fig.~\ref{fig:bound-states}. There are totally 194 bound states, including 12 
$s$-wave and 12 $p$-wave bound states. For more excited bound states with a 
smaller value of $|E_{n \ell}|$, there are more nodes in the wave function.

In the limit that $k_{n\ell} \, \Rsol \gg 1$, the bound state wave function outside the ball is exponentially small. In this limit, the bound state solution is well-approximated by the infinite well solution, for which $R_{n\ell}^{\rm out} \approx 0$. The energy eigenvalues are given by the condition $\kappa_{n\ell} \, = \, J_{\ell+1/2,n}/\Rsol$, where $J_{\ell+1/2,n}$ are the Bessel function zeroes. For $n \gg \ell$, these wavenumbers are well approximated by $\kappa_{n\ell} \, \approx \, [\pi (n + \ell/2 - 1/4)]/\Rsol$.

%%%%%%%%%%%%%%%%%%%%%%%%%%%%%%%%%%%%%
% Scattering States
%%%%%%%%%%%%%%%%%%%%%%%%%%%%%%%%%%%%%
\subsection{Scattering states}
\label{sec:Scattering-State}
%%%%%%%%%%%%%%%%%%%%%%%%%%%%%%%%%%%%%
The scattering state is also an eigenstate of energy, though with $E = k^2 / (2 \, \mu) > 0$.  It is \emph{not} an eigenstate of angular momentum as the scattering state is incident from far away with fixed momentum. The incident scattering wave function has the form $e^{i \, \mathbf{k} \cdot \mathbf{x}}$ far from the potential. As angular momentum is conserved in scattering off of a spherical potential, the full wave function can be decomposed as
\beqa
\psi_{\mathbf{k}}(\mathbf{x}) \, = \, \sum_{\ell m} R_{k \ell}(r) \, Y_{\ell m}^*(\hat{\mathbf{k}}) \, Y_{\ell m}(\hat{\mathbf{x}}) ~.
\eeqa
The radial wave function can also be expressed in terms of spherical Bessel 
functions,
\beqa
R_{k\ell}(r) = \left\{   \begin{array}{l l}  R_{k\ell}^{\rm in}(r) \, = \, c_3 \, j_\ell(\kappa \, r)\,,  &\quad r \leq \Rsol \,, \\ [1.0em]
    R_{k\ell}^{\rm out}(r) \, = \, c_1 \, j_\ell(k \, r) + c_2 \, y_\ell(k \, r) \,, &\quad  r > \Rsol  \,,
   \end{array}
\right.
\eeqa
where $\kappa \, = \, \sqrt{k^2 + 2 \, \mu \, V_0}$ and $k = \sqrt{2\,\mu\,E}$. 
Unitarity implies that scattering off of a spherical potential leads to a phase 
shift in the exterior wave function partial waves far away from the ball. In the 
decomposition of the wave function into partial waves, the phase shift condition 
corresponds to $\sqrt{c_1^2 + c_2^2} = 4 \pi i^\ell$. The remaining 
combinations 
of coefficients are determined by the boundary conditions $R_{k\ell}^{\rm 
in}(\Rsol) \, = \, R_{k\ell}^{\rm out}(\Rsol)$ and $R_{k\ell}^{\prime\,{\rm 
in}}(\Rsol) \, = \, R_{k\ell}^{\prime\,{\rm out}}(\Rsol)$. The coefficients for 
the exterior wave function can be written as
\beqa
c_1 \, = \, 4 \, \pi \, i^\ell \, \cos\delta \,, \qquad c_2 \,=\, 4 \, \pi \, i^\ell 
\,  \sin\delta \,,
\eeqa
where $\delta$ is the partial wave scattering phase, given by
\beqa
\label{eq:phase-shift}
\tan \delta \, = \, \frac{k \, j_\ell (\kappa \,\Rsol) \, j_{\ell+1}(k \, 
\Rsol) - \kappa \, j_{\ell+1}(\kappa \, \Rsol) \, j_\ell(k \, \Rsol)}{\kappa \, 
j_{\ell+1}(\kappa \, \Rsol) 
\, y_\ell(k \, \Rsol) - k \, j_\ell(\kappa \, \Rsol) \, y_{\ell+1}(k \, \Rsol)} ~.
\eeqa

For $s$-wave scattering state, the normalization factors have a simple analytic formula. For instance, 
\beqa
\label{eq:scattering-norm}
c_3 = \dfrac{4\sqrt{2}\,\pi\,\kappa}{\sqrt{\kappa^2 + k^2 + (\kappa^2 - k^2) \cos{(2\kappa\Rsol)}}} \approx \frac{4\pi}{|\cos{(\kappa \Rsol)}|}  ~,
\eeqa
where we have taken the limit of $k \ll \kappa$ for the small kinetic energy $E 
\ll V_0$. So, the amplitude of the inner-region wave function has a large peak 
value at $\Rsol = (2 m + 1) \pi/(2 \kappa)$ for integer $m$, which is 
coincident with the values of $\Rsol$ to have an $s$-wave resonant elastic 
scattering. In our later calculation of the radiative capture cross section, 
$\sigma_\gamma$, the oscillating peak structure of $c_3$ will induce a similar 
behavior for $\sigma_\gamma$ as a function of $\Rsol$. For other partial-wave 
scattering states, the similar resonant enhancement occurs for certain values 
of $\Rsol \approx J_{\ell-1/2,n}/\kappa$ where $J_{\nu,n}$ are Bessel function 
zeros.

\subsection{General scattering amplitude}
We begin by writing the general formula for the scattering amplitude. We then derive the analytic formulas to calculate the cross sections in two interesting limits: the dipole and low energy limits. 

The electromagnetic coupling of the nucleus to the vector potential is given by the interaction Hamiltonian
\begin{equation}\label{eq:interaction-H}
H_{\rm int} \,= \, \frac{1}{2 \, \mu} \, Z \, e \, \left[\mathbf{p}_{\rm N} \cdot \mathbf{A}(\mathbf{x}_{\rm N}) +  \mathbf{A}(\mathbf{x}_{\rm N}) \cdot \mathbf{p}_{\rm N}\right] ~,
\end{equation}
where $\mathbf{x}_{\rm N}$ and $\mathbf{p}_{\rm N}$ are the nucleus position 
and momentum operators respectively.  In the $\Msol \gg m_A$ limit, these 
nucleus operators reduce to $\mathbf{X} + \mathbf{x}$ and $\mathbf{p}$, where 
uppercase letters denote center-of-mass motion and lowercase letters denote 
relative motion. The scattering matrix element is then given by
\begin{equation}\label{eq:gen-amplitude}
\mathcal{M}_{n\ell m} \, = \, \frac{1}{2 \, \mu} \, Z \, e \,  \boldsymbol{\epsilon}^* \cdot \int d^3 x \, e^{-i \, \mathbf{q} \cdot \mathbf{x}} \left[\nabla \psi_{n\ell m}^*(\mathbf{x})\, \psi_{\mathbf{k}}(\mathbf{x}) - \psi_{n\ell m}^*(\mathbf{x})\, \nabla\psi_{\mathbf{k}}(\mathbf{x})\right] ~,
\end{equation}
where $\boldsymbol{\epsilon}=\boldsymbol{\epsilon}(\mathbf{q})$ is the photon 
polarization satisfying $\mathbf{q} \cdot \boldsymbol{\epsilon}(\mathbf{q}) = 
0$ and $\psi_{\mathbf{k}}$/$\psi_{n\ell m}$ are the scattering/bound state 
wave functions relative to the center of mass respectively. Note that the 
scattering and bound state wave functions have different normalizations and 
different mass dimensions. The scattering wave function is not normalizable as 
the incident wave is a plane wave. The photon momentum and  energy have 
$|\mathbf{q}|=\omega_{n\ell} \approx  E_{\mathbf{k}} + |E_{n \ell}|$. For a 
small dark matter velocity, the kinetic energy is in general smaller than the 
binding energy and the photon energy is approximately the binding energy.

The radiative capture cross section in the non-relativistic normalization is then given by
\beqa
\sigma_{\gamma, n\ell} \, =\, \frac{1}{v}\, \int d\Omega\,\frac{|E_{n\ell}|}{8 \, \pi^2} \,\sum_m\, |\mathcal{M}_{n\ell m}|^2 ~.
\eeqa
In general, one can keep all partial wave functions in the scattering state, 
expanding $e^{-i\,\mathbf{q}\cdot\mathbf{x}}$ in partial waves as well and 
performing the integration to calculate $\sigma_{\gamma, n\ell}$. This 
procedure is conceptually clear, but practically tedious. Instead, we mainly 
focus on few parameter regions with good approximation schemes, derive analytic 
formulas and present the cross sections based on them. 

The most relevant parameters for radiative capture of the \EDB/ is the radius $\Rsol$, the scattering kinetic energy or momentum $k \equiv |\mathbf{k}|$ and the radiated photon energy $q \equiv |\mathbf{q}|$. The three limits are
\begin{itemize}
\item {\it Dipole limit}:  $q \Rsol \ll 1$. In this limit, the wavelength of the emitted photon is much larger than the radius of the DMB, so that the wave function of the emitted photon becomes trivial or $e^{-i\,\mathbf{q}\cdot\mathbf{x}} \rightarrow 1$.
\item {\it Low energy limit}: $k  \Rsol \ll 1$. In this limit, only the $s$-wave mode of the scattering state has a significant contribution. 
\item {\it Semi-classical limit}: $q \Rsol \gg 1$ and $k \Rsol \gg 1$. In this limit, a large number of closely spaced bound states can be produced and all wave functions are oscillating rapidly across the potential. 
\end{itemize}
For the semi-classical limit, we cannot take the classical limit as the dominant radiation effect comes from the suppressed quantum effects over the entire radius of the potential. This limit is the most challenging to compute due to the large number of contributing states and amplitudes.  We approximate the cross section via scaling relations inferred from the other two limits as well as the behavior of the neutron capture cross sections by a nucleus in the similar limit. 

We also note that the hydrogen radiative capture process satisfies both dipole 
limit and low energy limit because of the smallness of the electromagnetic 
coupling and the long-range property of Coulomb interactions. A simple analytic 
result can therefore be obtained in that 
case~\cite{Kramers,1978JETP...48..639K,An:2016gad}. 

\subsection{Dipole Limit}
\label{section:dipole}

In the dipole limit with $q \Rsol \ll 1$, $|\mathbf{q}\cdot \mathbf{x}| \ll 1$ for all $|\mathbf{x}| \leq \Rsol$.  In this limit, we simplify the calculation using the operator relation $\mathbf{p} \, = \, -i \, \mu \, [\mathbf{x},H]$~\cite{Sakurai:1167961}, where $H$ is the Hamiltonian excluding the interaction \eqref{eq:interaction-H}.  The matrix element becomes
\beqa
\label{eq:dipole-amplitude}
|\mathcal{M}_{n\ell m}| \, = \, Z \, e \, (|E_{n\ell}| +E_{\mathbf{k}}) 
\, \int d^3x \, \psi_{n \ell m}^*(\mathbf{x}) \, \boldsymbol{\epsilon}^* \cdot \mathbf{x} \, \psi_{\mathbf{k}}(\mathbf{x}) ~.
\eeqa
Starting with Eq.~\eqref{eq:dipole-amplitude}, we perform the angular integration over spherical harmonics, square the amplitude, integrate over the photon emission angle, and sum over photon polarizations and $m$ of the bound states to obtain the radiative capture cross section for the bound state $(n, \ell)$
\beqa
\label{eq:dipole-xsec}
\sigma_{\gamma, n\ell} &=& \frac{1}{v}\,\frac{Z^2 \, \alpha}{3 \, \pi} \, |E_{n\ell}|(|E_{n\ell}| + E_{\mathbf{k}})^2 \,   \nonumber \\
&&  \quad\times \left[\ell\, \left|\int dr \, r^3 \, R_{n\ell}(r) \, R_{k\ell-1}(r) \right|^2 + (\ell+1)\, \left|\int dr \, r^3 \, R_{n\ell}(r) \, R_{k\ell+1}(r) \right|^2\right] ~.
\eeqa

\begin{figure}[thb!]
	\centering
	\includegraphics[width=0.6\textwidth]{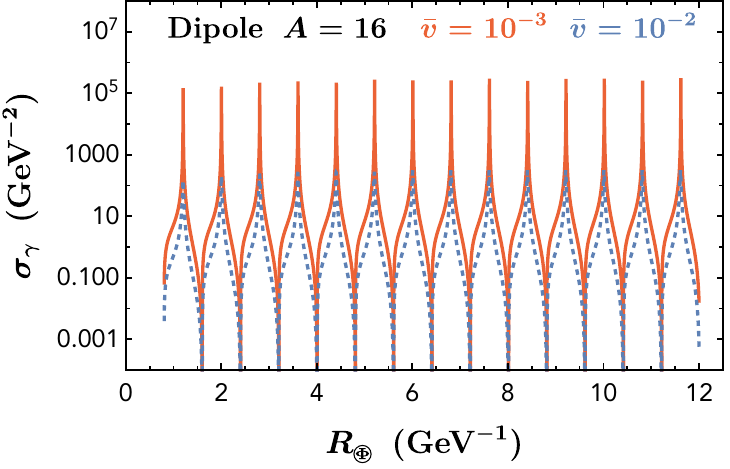}
	\caption{Radiative capture cross section as a function of the \DMB/ radius 
	for the oxygen nucleus with $Z=8$ and $A = 16$ in the dipole approximation. 
	The dominant $p$-wave bound states and $s$-wave scattering states are 
	included in this plot. Two different dark matter averaged velocities of 
	$\bar{v} = 10^{-3}$ and $\bar{v} = 10^{-2}$ are considered.}
	\label{fig:radial-dep-dipole}
\end{figure}

Using the oxygen nucleus as an example, we show the radiative capture cross 
section as a function of $\Rsol$ in Fig.~\ref{fig:radial-dep-dipole}. Since 
photon energy can be as large as the depth of potential barrier, $V_0 \approx A 
\times 32$~MeV, and if all deep bound states are included, the dipole 
approximation is only valid up to a radius of $\Rsol \approx 2\pi/V_0 \approx 
12$~GeV, which sets the upper end of the $x-$axis in Fig.~\ref{fig:radial-dep-dipole}. The 
oscillating peak structure is also obvious for this plot. As we discussed 
around \eqref{eq:scattering-norm}, the scattering states could be close to a 
bound state for a certain $\Rsol$, leading to resonant enhancement of the cross 
section.

\subsection{Low Energy Limit}
\label{sec:low-energy}

In the low energy limit with $k  \Rsol \ll 1$, we find it convenient to use integration by parts and the on-shell photon conditions $\mathbf{q} \cdot \boldsymbol{\epsilon} = 0$ to rewrite the amplitude as
\beqa
\label{eq:swave-amplitude}
\mathcal{M}_{n\ell m} \, = \, -\frac{1}{\mu} \, Z \, e \,  \boldsymbol{\epsilon}^* \cdot \int d^3 x \, e^{-i \, \mathbf{q} \cdot \mathbf{x}} \, \psi_{n\ell m}^*(\mathbf{x})\, \nabla\psi_{\mathbf{k}}(\mathbf{x}) ~.
\eeqa
For the dipole factor $e^{- i \, \mathbf{q} \cdot \mathbf{x}}$, we decompose its complex conjugate in spherical harmonics as
\beqa
e^{ i \, \mathbf{q} \cdot \mathbf{x}} \, = \, \sum_{\ell^\prime,m^\prime} 4 \, \pi \, i^{\ell^\prime} \, j_{\ell^\prime}(q \, r) \, Y_{\ell^\prime m^\prime}^*(\hat{\mathbf{q}}) \, Y_{\ell^\prime m^\prime}(\hat{\mathbf{x}}) ~.
\eeqa
The scattering state wave function for $k r \, \ll \, 1$ outside the \DMB/ scales as
\beqa
\psi_{\mathbf{k}} \, \approx \, \sum_{\ell m} a_{k \ell}\, (k  r)^\ell \, Y_{\ell m}^*(\hat{\mathbf{k}}) \, Y_{\ell m}(\hat{\mathbf{x}}) \approx a_{k0} \, Y_{00}^*(\hat{\mathbf{k}}) \, Y_{00}(\hat{\mathbf{x}}) ~,
\eeqa
where $a_{k\ell}$ are non-zero numerical coefficients.  In other words, for $k \Rsol \ll 1$, the $s$-wave term dominates at the boundary of the potential, which can further simplify our calculation.  Putting these pieces together, squaring, summing over polarizations and the final state $m$ number and integrating over the photon emission angle, we find the cross section is given by
\beqa
\label{eq:lowe-xsec}
\sigma_{\gamma, n\ell} \, = \, \frac{1}{v} \, \ell \, (\ell + 1)\,(2 \ell + 1) \,\frac{Z^2 \, \alpha \, |E_{n\ell}|}{2 \, \pi \, \mu^2 \, q^2} \,  \left|\int dr \, r \, j_\ell(q \, r)\, R_{n\ell}(r) \, R_{k0}^\prime(r)\right|^2 \qquad (\ell \geq 1) ~.
\eeqa
\begin{figure}[thb!]
\centering
\includegraphics[width=0.48\textwidth]{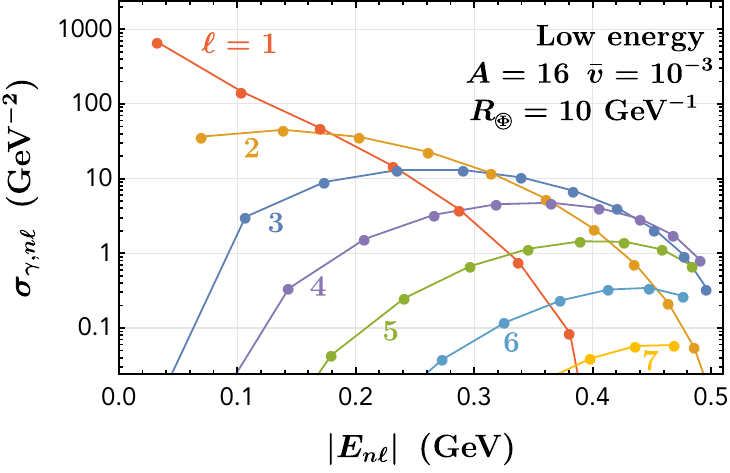} \hspace{3mm}
\includegraphics[width=0.485\textwidth]{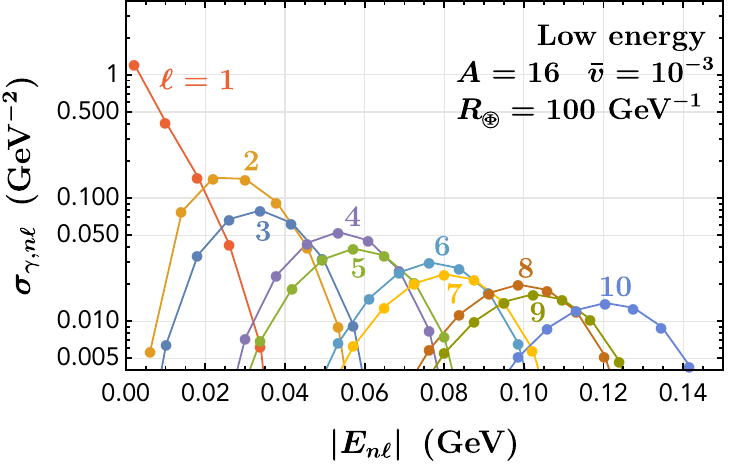}
\caption{Radiative capture cross section as a function of the binding energy of 
the bound states for $Z = 8$, $A = 16$, $V_0 = A \times 32~{\rm MeV}$ and 
$\bar{v} = 10^{-3}$, as well as $\Rsol = 10~{\rm GeV}^{-1}$ (left) and $\Rsol = 
100~{\rm GeV}^{-1}$ (right).}\label{fig:lowe-Elevels} 
\end{figure}

We first note that the above formula could have a broader application beyond 
the dark matter phenomenology here. The radiative capture cross section in the 
low energy limit could also be applied to other quantum mechanical systems. 
Secondly, note that in the simultaneous dipole and low energy limits, both 
\eqref{eq:dipole-xsec} and \eqref{eq:lowe-xsec} reduce to~\footnote{For 
Eq.~\eqref{eq:lowe-xsec} in the low-energy limit, one can use integration by 
parts and the Schr\"odinger equations to derive \eqref{eq:common-formula}.}
\beqa
\label{eq:common-formula}
\sigma_{\gamma, n1} \, = \,\frac{1}{v}\, \frac{Z^2 \, \alpha}{3 \, \pi} |E_{n1}|\, (|E_{n1}| + E_{\mathbf{k}})^2 \left|\int dr \, r^3 \, R_{n1}(r) \, R_{k0}(r)\right|^2 ~. 
\eeqa
In this simultaneous limit, an analytic result for the cross section can be 
obtained,
\beqa
\sigma_{\gamma, n1} & = & \frac{1}{v}\, \frac{Z^2 \, \alpha}{3 \, \pi \, 
N_{n1}^2} \, \Rsol^8 \, |E_{n1}|\, (|E_{n1}| + E_{\mathbf{k}})^2
 \times  \\
& &\hspace{-15mm} \left\{ \frac{c_3\,\Big\{\hat{\kappa} \, \cos\hat{\kappa} \, [(\hat{\kappa}^2 - 3 \, \hat{\kappa}_{n1}^2) \, \sin\hat{\kappa}_{n1} +  \hat{\kappa}_{n1} \, 
\Delta\hat{\kappa}^2 \, \cos\hat{\kappa}_{n1}] +\hat{\kappa}_{n1}^2 \, \sin\hat{\kappa} \, ( \Delta\hat{\kappa}^2 \, \sin\hat{\kappa}_{n1} + 2\, \hat{\kappa}_{n1} 
\,\cos\hat{\kappa}_{n1}) \Big\}  }  { \hat{\kappa}\, (\Delta\hat{\kappa}^2)^2 \, 
(\hat{\kappa}_{n1} \, \cos\hat{\kappa}_{n1} - \sin\hat{\kappa}_{n1})} 
\right. \nonumber\\ 
& &  \left. \hspace{-10mm} + \frac{4 \, \pi  \Big\{\hat{k} \, [\hat{k}^2 \, (1 + \hat{k}_{n1}) + \hat{k}_{n1}^2 
\, (3 + \hat{k}_{n1})] \, 
\cos(\hat{k} - \delta) + \hat{k}_{n1}^2 (\hat{k}^2 + 
\hat{k}_{n1}^2 + 2 \, \hat{k}_{n1}) \, 
\sin(\hat{k} - \delta)\Big\}  }{\hat{k} \, (\hat{k}^2 + 
\hat{k}_{n1}^2)^2 \, (\hat{k}_{n1} + 1)}  \right\}^2,\nonumber
\eeqa
where $\delta$ is the elastic scattering phase, given in our convention by $\tan \delta = c_2 / c_1$ from Eq.~\eqref{eq:phase-shift}, $\hat{x} = x \, \Rsol$, and $\Delta \hat{\kappa}^2 = 
\hat{\kappa}_{n1}^2 - \hat{\kappa}^2$. The factor $c_3$ is given in \eqref{eq:scattering-norm}, while $N_{n1}^2$ given in \eqref{eq:bound-norm}.

After numerical integration of Eq.~\eqref{eq:lowe-xsec}, we show the cross sections for  two benchmark radii in Fig.~\ref{fig:lowe-Elevels} for illustration, $\Rsol = 10~{\rm GeV}^{-1}$ and $\Rsol = 100~{\rm GeV}^{-1}$, along with the oxygen element and $\overline{v} = 10^{-3}$. For $\Rsol = 10~{\rm GeV}^{-1}$, the energy levels can be found in the left panel of Fig.~\ref{fig:bound-states}, while the $\Rsol = 100~{\rm GeV}^{-1}$ case shows similar behaviors. The most excited $p$-wave state has $E_{11} \approx 32.7$~MeV for $\Rsol = 10~{\rm GeV}^{-1}$ and $E_{11} \approx 2.2$~MeV for $\Rsol = 100~{\rm GeV}^{-1}$. As can be seen from Fig.~\ref{fig:lowe-Elevels}, 
the radiative capture cross section is dominated by the most excited state of 
$\ell=1$ bound states, which is precisely the limit in which the dipole 
approximation applies. For a fixed $\ell$, the cross section decreases 
exponentially for deeper bound states with larger binding energy. 

\begin{figure}[t!]
	\centering
	\includegraphics[width=0.48\textwidth]{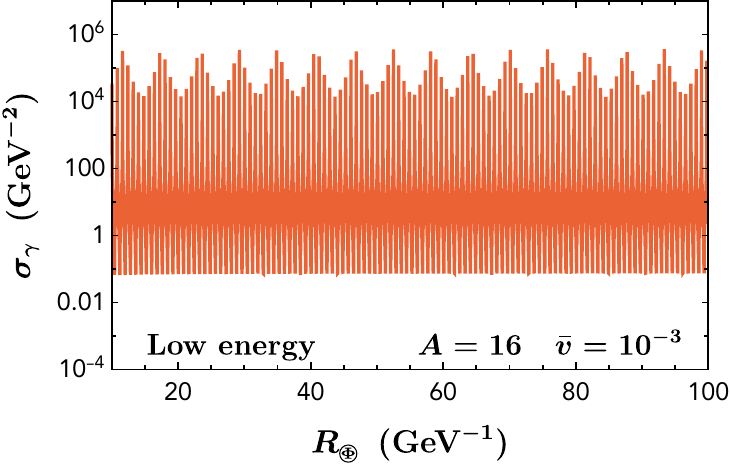} \hspace{3mm}
	\includegraphics[width=0.48\textwidth]{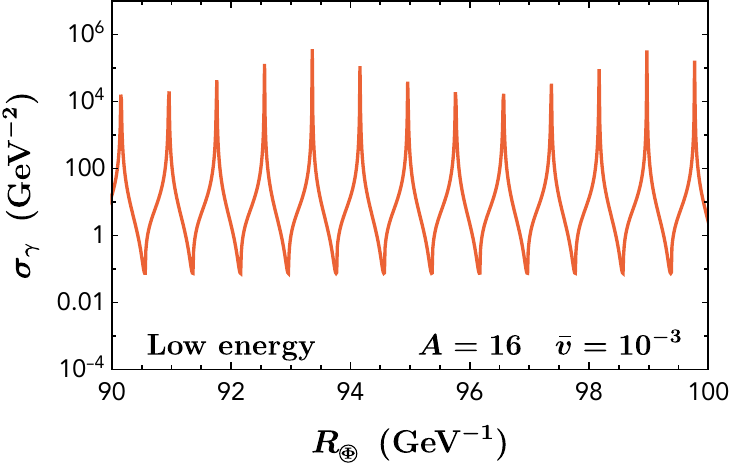}
	\caption{Radiative capture cross section as a function of the \DMB/ radius for $Z=8$, $A = 16$, $V_0 = A \times 32~{\rm MeV}$, and $\bar{v} = 10^{-3}$ in the low energy limit, in which only $s$-wave scattering state is included. The right panel narrows the range to the largest radii considered.}\label{fig:radial-dep-lowe}
\end{figure}

In Fig.~\ref{fig:radial-dep-lowe}, we show the capture cross section as a 
function of $\Rsol$ up to a radius slightly smaller than $2\pi/k \approx 
2\pi/(A\,m_p\,\bar{v}) \approx 400~{\rm GeV}^{-1}$. Again, one can see a clear 
oscillation behavior, which is due to the resonance effects in the scattering 
state. The cross section envelope has a mild dependence on the radius, 
although it is very sensitive to the actual value of $\Rsol$ within one period 
of the wave.

\begin{figure}[thb!]
	\centering
	\includegraphics[width=0.6\textwidth]{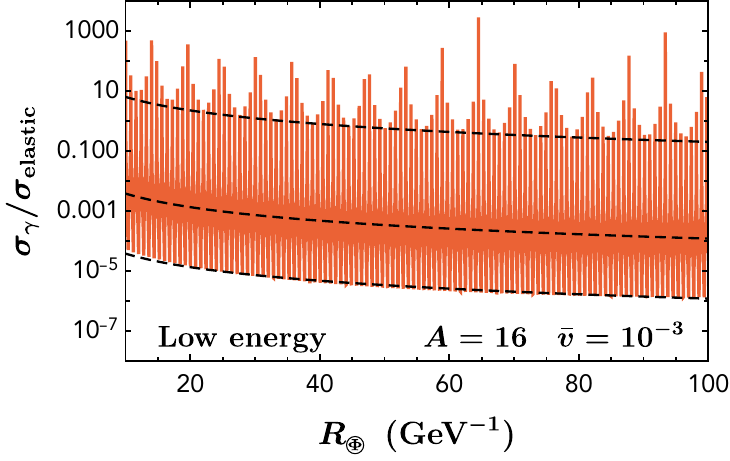}	
	\caption{Ratio of the radiative capture to elastic scattering cross section of \DMB/ in the low energy limit. The benchmark $Z=8$, $A = 16$, $V_0 = A \times 32~{\rm MeV}$ and $\bar{v} = 10^{-3}$ is used. The dashed and black line has the ratio $\propto \Rsol^{-3/2}$.}\label{fig:elastic-cap}
\end{figure}

It is instructive to compare the radiative capture cross section to the elastic scattering cross section. Using the phase shift method, the elastic scattering cross section is calculated in the low energy limit by
\beqa
\sigma_{\rm elastic} \approx \frac{4 \, \pi}{\kappa^2} \, \left[\tan{(\kappa  \Rsol)} - \kappa \Rsol\right]^2 ~,
\eeqa
which has a similar oscillating behavior with the same periodicity. The ratio 
of the radiative capture cross section (in the region under computational 
control) to this value is shown in Fig.~\ref{fig:elastic-cap}, which still has 
an oscillating behavior. In the dashed and black lines, we guide the general 
envelop behavior of this ratio. The general behavior of this ratio as a 
function of $v$ and $\Rsol$ has a simple scaling 
\beqa
\sigma_\gamma/\sigma_{\rm elastic} \propto v^{-1}\, \Rsol^{-3/2}  ~,
\eeqa
with the range of radii satisfying the low energy approximation. 

Note that as $\Rsol$ is varied, the scattering wave function in each partial wave mode can be resonantly enhanced when
\begin{equation}
\Rsol \approx \frac{J_{\ell-1/2,n}}{\kappa} ~,
\end{equation}
where $J_{\nu,n}$ are Bessel function zeros. The $s$-wave wave function gets 
enhanced by a factor of $\kappa / k$, leading to an enhancement of the 
radiative capture cross section by $(\kappa / k)^2$.                         

The dependence on the \DMB/ radius of the total radiative capture cross section 
in the dipole limit is shown in Fig.~\ref{fig:radial-dep-dipole}. Beyond $\Rsol 
\approx 12~{\rm GeV}^{-1}$, we work in the low energy approximation.  The 
dependence on $\Rsol$ in this limit is shown in 
Fig.~\ref{fig:radial-dep-lowe}.  Beyond $\Rsol \approx 100~{\rm GeV}^{-1}$, the 
low energy limit is no longer applicable.

\subsection{Large Radius Limit}
\label{sec:large-radius}

In the large radius limit, the approximations we have used cease to apply and 
the calculation of the radiative capture cross section becomes computationally 
prohibitive. The elastic scattering cross section, on the other hand, can be 
determined in this limit by summing our analytic expression for the partial 
wave cross section to a sufficiently high partial wave number. As seen in Ref.\ 
\cite{Ponton:2019hux}, it saturates the geometric cross section $\pi \Rsol^2$ 
up to an 
$\mathcal{O}(1)$ factor. We proceed by estimating the ratio of the radiative 
capture cross section to the known elastic cross section in two different ways: 
by extrapolating the ratio shown in Fig.~\ref{fig:elastic-cap} to large radius 
and by determining this ratio in neutron capture data. Neither procedure is 
entirely robust, but they are meant to provide a guideline for the 
possibilities.

In the low energy limit, we have found that the ratio of the radiative capture 
to elastic scattering cross sections scales as $\Rsol^{-3/2}$.  Extrapolating this 
behavior to large $\Rsol$ indicates that the radiative capture cross section 
scales as $\Rsol^{1/2}$.  We estimate that for $k \Rsol \gg1$ the $\sigma_\gamma$ should saturate 
to
\beqa
\sigma_\gamma \sim 60 \,{\rm GeV}^{-2} \times \left(\frac{10^{-3}}{v}\right)\, \left(\frac{\Rsol}{10^5~{\rm 
GeV}^{-1}} \right)^{1/2}~.
\eeqa

\begin{figure}[thb!]
	\centering
	\includegraphics[width=0.6\textwidth]{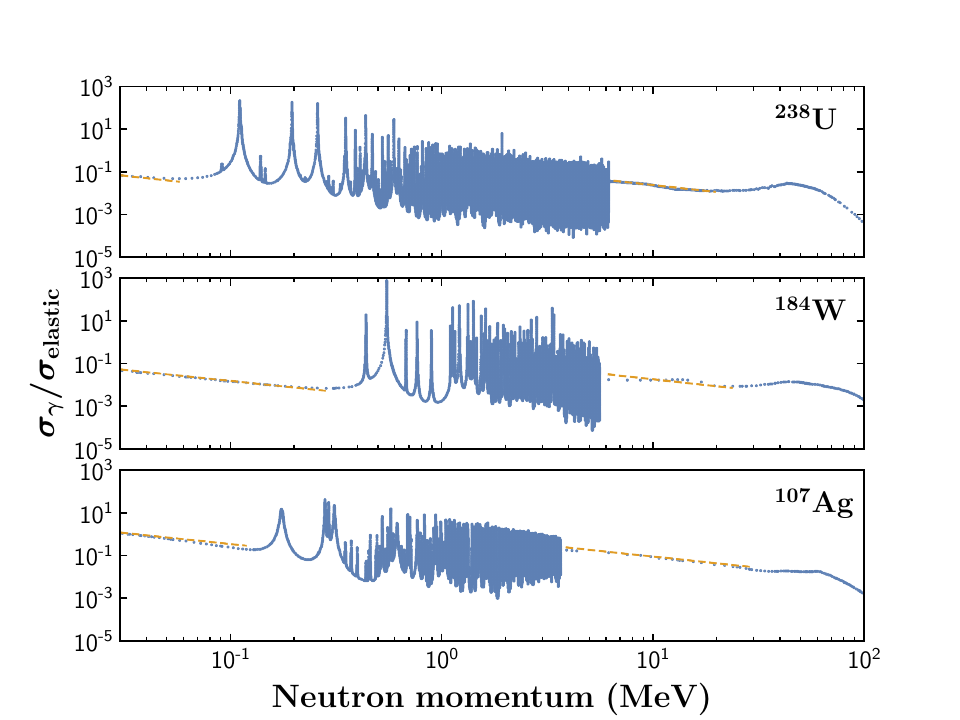}
	\caption{Ratio of the radiative capture to elastic cross sections for 
	neutron capture by three different nuclei using the TENDL-2017 
	model~\cite{osti_22131266,2017EPJWC.14607006S} in blue color. The dashed 
	orange lines indicate a $1/v$ behavior expected at energies below the 
	nuclear resonances and at energies above the nuclear resonances, but below 
	onset of free nucleon scattering.}\label{fig:nuclear}
\end{figure}
Alternatively, this ratio can be estimated from neutron radiative capture data (see Fig.~\ref{fig:nuclear}). 
The data are plotted as a function of the incident neutron momentum. 
Since the relevant comparison for determining the large $\Rsol$ limit is $k 
\Rsol 
\gg 1$, the limit is expected to be reached at large kinetic energy when the 
momentum becomes comparable to the effective inverse radius of the nucleus. 
Only for relatively heavy elements does radiative capture reach the large $k 
\Rsol$ 
limit below the complicated MeV scale. The data are not sufficiently 
homogeneous across different nuclei to determine a clear numerical pattern.  
Nevertheless, the ratios are seen to follow the expected qualitative behavior, 
going to a smooth function in the large $k \Rsol$ limit (the region to the 
right of 
the resonance region). The ratios of radiative 
capture to elastic scattering cross sections for isotopes of uranium, tungsten and silver
are shown in Fig.~\ref{fig:nuclear} using the TENDL-2017 
model~\cite{osti_22131266,2017EPJWC.14607006S}. These nuclei are chosen as 
cases where there is a significant amount of data that agree with the 
model, as the quality and availability of data varies widely between 
different isotopes.

\section{Prospects for detection}
\label{sec:detection}

Radiative capture of nuclei by MDM entering the detector deposits significantly 
more energy than elastic scattering. The radiation from the initial capture is 
seen in Fig.\ \ref{fig:lowe-Elevels} to be of order MeV or larger. Furthermore, 
excited states are typically produced; their subsequent decay leads to 
additional emission totaling around 100s of MeV. 

A full study of these signals in individual detectors is beyond the scope of 
this work. Nevertheless, we consider some basic properties of current and 
forthcoming detectors to determine the viability of this signal. Direct 
detection experiments such as Xenon1T~\cite{Aprile:2017aty} and 
LZ~\cite{Akerib:2019fml} should be sensitive to radiative capture as the 
deposited energy far exceeds their threshold. Current analyses veto multiple energy 
deposits in a short time window~\cite{Aprile:2019bbb}. A single energy deposit 
would likely be beyond their current search window, but should be visible if it 
exceeds the radioactive background. More striking would be several deposits of 
comparable energies. We consider this multi-hit signal as our primary signal.

The main advantage of considering radiative capture signals, however, is to 
consider higher energy threshold, but larger detectors. These detectors are primarily 
large neutrino detectors. The threshold in the full IceCube volume is at the 
100 GeV scale~\cite{deWasseige:2019xcl}, above the energies typically deposited by MDM. The next largest that 
could be sensitive is Hyper-Kamiokande~\cite{Abe:2014oxa}. A few MeV energy 
deposit, which is the energy released in the initial capture, is close to the 
threshold and, even if it is reconstructable,  may not 
be distinguishable from radioactive and other backgrounds if it is isolated. We 
therefore put the requirement that at least 5 capture events
occur during MDM passage through the detector for all detectors considered for 
our main analysis. This
could be particularly striking at detectors with tracking capabilities like 
DUNE~\cite{Acciarri:2015uup}, where the deposits would form a line in the 
detector.  Highly segmented detectors like NO$\nu$A~\cite{Ayres:2007tu} may 
also be able to track the MDM passage in 2D.

A multi-hit signal would be a spectacular event that would be hard to fake with 
any background. It may, however, not be required in order to identify an event 
with radiative capture during MDM passage. The radiative capture event 
typically 
produces a highly excited bound state, which de-excites and leads to further 
photon production. This proceeds until the ground state is reached with a 
binding energy of 100s of MeV, that is with the release of 100s of MeV in 
photons. Such photons are unlike the dominant potential backgrounds from solar 
and atmospheric neutrinos in their energy spectrum and topology respectively, 
while they are above the radioactive decay energy range. Cosmic rays could be 
an additional background for surface detectors like 
ProtoDUNE~\cite{Abi:2017aow}, though the 
cosmic ray tagger should reduce this background significantly. If the search 
can indeed be made background free by judicious selection criteria, then a very 
small number of radiative capture single-hit events would be required for a 
discovery.  The expected event rate scales like $1/\Msol$ until a point is 
reached that the number of MDM passing through the detector over the course of 
the experiment running time is less than one. At that point, 
corresponding 
to an upper bound on $\Msol$, expected sensitivity is lost. A detailed study 
of the feasibility of such a low event count search is beyond the scope of this 
work.

We now comment further on the capabilities of water Cherenkov, liquid argon time projection chamber (LArTPC), and 
liquid scintillator detectors to detect photons of the relevant energies. In 
any of these detectors, the visible result of the photon that is 
emitted during radiative capture is an electron/positron produced by Compton 
scattering at low energies or $e^+e^-$ pair production above 1 MeV. The 
remaining question is what is the threshold for detecting the charged particles 
produced in these processes.

The largest suitable water Cherenkov detector at the moment is 
Super-Kamiokande~\cite{Fukuda:2002uc}. 
Another detector, Hyper-Kamikande, with similar 
technology, but an order of magnitude larger mass and volume is planned. The 
physical 
threshold is given by the Cherenkov momentum of the electron in water, namely 
$p = 583~{\rm keV}$. In principle, 
any photon above the corresponding Cherenkov 
kinetic energy of $263~{\rm keV}$ can produce a visible electron. The 
efficiency for detection at these low energies is likely to be poor, but if 
multiple such events are lined up through the detector, the detection prospects 
may be improved. In practice, the lowest threshold analyses have pushed down to 
electron kinetic energies of $3.5~{\rm MeV}$~\cite{Sekiya:2016nnn}. 

LArTPC is an up-and-coming detector technology. 
The largest such detector will be the Deep Underground Neutrino Experiment 
(DUNE). Other smaller detectors based on this technology, such as 
ProtoDUNE~\cite{Abi:2017aow}, 
are currently operating. In between sits the ICARUS 
experiment~\cite{Antonello:2015lea}, slated to start taking data in December 
2019. These detectors track charged particles as they pass through the Argon. 
The threshold energy at DUNE and in LArTPC detectors in general is less well known as reconstruction in these detectors is still under active development. The limit 
is set by the travel distance of the electron in liquid Argon. The TPC wire 
pitch is roughly 4 mm at DUNE, so the electron should travel at least that 
distance to have multiple hits, corresponding to a kinetic energy of around 900 
keV. In order to reconstruct particles, at least 10 hits are generally 
required. In other words, the electron should travel at least 40 mm, 
corresponding to a kinetic energy of around 9 MeV. On the other hand, the DUNE 
CDR sets a $e^{\pm}/\gamma$ threshold at 30 MeV~\cite{Acciarri:2015uup}, as low 
energy electromagnetic particles may become difficult to disentangle from other 
charged particles such as muons. Above 30 MeV, an electron will typically 
undergo Bremsstrahlung before stopping, leading to a characteristic shower.

The largest liquid scintillator detectors are currently 
Borexino~\cite{Alimonti:2008gc} and the NO$\nu$A~\cite{Ayres:2007tu} far 
detector. JUNO~\cite{An:2015jdp}, a forthcoming detector 
that also uses a liquid scintillator, will be significantly larger. The medium 
in the latter is a linear alkyl benzene (LAB) that is planned to have excellent 
energy resolution. 
The LAB is 88\% carbon by mass. The low energy reconstruction is limited by the 
requirement of a minimum number of photoelectrons to reconstruct the radiative 
capture event. The threshold for the trigger can be as low as 0.065 
MeV~\cite{Shah:2018} and easily below 0.5~MeV~\cite{Fang:2019lej}, which should 
easily be able to detect a radiative capture 
event. Borexino has a similarly low 
threshold~\cite{Alimonti:2008gc}. The NO$\nu$A detector is designed for higher 
energy events and a threshold around 15 MeV~\cite{Ayres:2007tu}. This would 
likely make it challenging to observe the capture photon, but the de-excitation 
photons could still be seen.

\begin{table}[!htb]
	\centering
	\begin{tabular}{l@{\hspace{0.7cm}}c@{\hspace{0.6cm}}c@{\hspace{0.6cm}}c@{\hspace{0.6cm}}c}
		\hline\hline
		Detector & $A_{\rm eff}~({\rm cm}^2)$ & 
		$L_{\rm eff}~({\rm cm})$ & Nuclei & $n_A~(10^{22}~{\rm cm}^{-3})$ \\
		\hline
		Xenon 1T & $1.09 \times 10^5$ & 64.2 & Xe & $1.42$ \\
		LZ & $2.65 \times 10^5$ & 100 & Xe & $1.42$ \\
		ICARUS & $5.08 \times 10^5$ & 255 & Ar & $2.10$ \\
		ProtoDUNE & $6.78 \times 10^5$ & 446 & Ar & $2.10$ \\
		Borexino & $5.67 \times 10^5$ & 567 & C & $3.96$ \\
		No$\nu$A & $1.17 \times 10^7$ & 1400 & C & $3.10$ \\
		JUNO & $9.84 \times 10^6$ & 2360 & C & $3.79$\\
		Super-Kamiokande & $1.88 \times 10^7$ & 2670 & O & $3.34$\\
		DUNE & $2.14 \times 10^7$ & 2290 & Ar & $2.10$\\
		Hyper-Kamiokande & $1.12 \times 10^8$ & 4580 & O & $3.34$\\
		\hline\hline
	\end{tabular}
	\caption{Properties of the detectors considered in our estimation of MDM 
	discovery potential.}\label{tab:detectors}	
\end{table}
The properties of the detectors, including geometry and the largest nuclei that 
make up a significant fraction of the medium are also provided in Table 
\ref{tab:detectors}. The 
effective area $A_{\rm eff}$ of the detector is the average area normal to the 
DM trajectory over the DM velocity distribution and uniform position 
distribution. The effective 
length $L_{\rm eff}$ is the average length of the DM path through the detector. 
For detectors with multiple modules, it is assumed that the modules are 
sufficiently closely spaced that they operate functionally as a single large 
detector. 
Radiative capture is dominated by the most massive common nuclei in the 
detector, so we consider only interactions with these nuclei. The number 
density of the dominant nuclei are denoted by $n_A$. 

\begin{figure}[t!]
	\centering
	\includegraphics[width=0.7\textwidth]{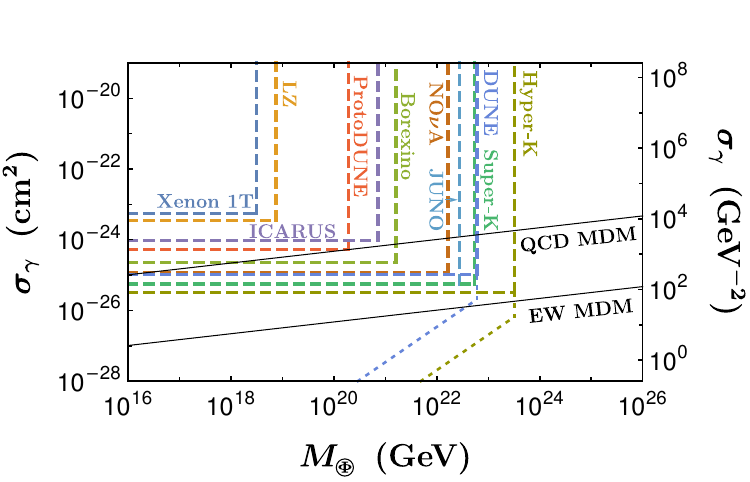}
	\caption{Sensitivity of Xenon 1T (blue)~\cite{Aprile:2017aty}, LZ 
	(orange)~\cite{Akerib:2019fml}, ProtoDUNE (red)~\cite{Abi:2017aow},
	ICARUS (purple)~\cite{Antonello:2015lea}, Borexino 
	(green)~\cite{Alimonti:2008gc}, NO$\nu$A (brown)~\cite{Ayres:2007tu}, JUNO 
	(light blue)~\cite{An:2015jdp}, 
	Super-Kamiokande (teal)~\cite{Fukuda:2002uc}, DUNE 
	(dark blue)~\cite{Acciarri:2016crz}, 
	and  Hyper-Kamiokande (dark green)~\cite{Abe:2014oxa} to radiative capture 
	of nuclei by MDM. The experiments are listed in order of increasing mass 
	sensitivity. The dashed lines indicate the 
	sensitivity if at least 5 radiative capture events are required for each 
	MDM passage through the detector. A running time of 10 years is assumed for 
	Borexino, Super-Kamiokande, DUNE, and Hyper-Kamiokande, 5 years is assumed 
	for ICARUS and NO$\nu$A and one year at the direct detection experiments 
	and ProtoDUNE. The dotted lines for DUNE and Hyper-Kamiokande indicate the 
	region in which at least one radiative capture event is expected over the 
	whole running time. The black line indicates the estimated radiative 
	capture cross section for QCD and EW density MDM.}\label{fig:limits}
\end{figure}
Given this analysis strategy, we proceed to determine the region of parameter 
space to which each of these detectors is sensitive. We parameterize 
the models in terms of the MDM mass and radiative capture cross section of 
the heaviest nucleus in the detector in question in order to maintain model 
independence. Pending a detailed study, we assume that 5 energy deposits during 
the MDM passage is reconstructed with 100\% efficiency and is efficiently 
separable from potential backgrounds such as radioactive decays, cosmic rays 
and neutrinos. The resulting 
estimated sensitivity is presented in Fig.~\ref{fig:limits}. We assume a one 
year running time for Xenon 1T, LZ, and ProtoDUNE, a 5 year running time at 
ICARUS and NO$\nu$A and a 10 year running time for the remaining large neutrino 
experiments. In addition, for the DUNE and Hyper-Kamiokande experiments, we 
indicate the region in which at least one radiative capture event will occur 
over 10 
year in order to assess the potential sensitivity of a single-hit 
analysis.

%%%%%%%%%%%%%%%%%%%%%%%%%%%%%%%%%%%%%
% Conclusion
%%%%%%%%%%%%%%%%%%%%%%%%%%%%%%%%%%%%%
\section{Discussion and conclusions}
\label{sec:conclusion}
%%%%%%%%%%%%%%%%%%%%%%%%%%%%%%%%%%%%%
The signal of radiative capture appears to be rather striking and should be 
detectable by any of the experiments considered in this work. That said, a full 
study of the reconstruction of this signal in each experiment is left to future 
work. In particular, we have left open the question of whether single radiative 
capture event can be reconstructed and distinguished from background or 
whether even more distinctive multi-hit events are required in order to have a 
background free search. Furthermore, triggering and data acquisition may be an 
issue in large, high granularity detectors. While it should be possible to 
design a system to collect radiative capture events, the current systems may 
unfortunately miss these events. We also note that searching for radiative 
capture events is complimentary to performing traditional dark matter direct 
detection experiments, which are searching for single hit, low energy threshold 
elastic scattering events. For a large radius, above the threshold to have a 
bound 
state, radiative capture events have an advantage because one can utilize a 
larger detector with a higher threshold. For a small radius, traditional 
direct detection experiments have an advantage due to their a low energy 
threshold.

In summary, we have studied the process of radiative capture nuclei by 
macroscopic dark matter. We have found that the radiative capture process can 
be comparable to the elastic scattering process, but has generally more 
promising detection prospects at higher energy threshold neutrino detectors. Current 
large neutrino detectors such as ProtoDUNE,  ICARUS,  Borexino, NO$\nu$A and 
Super-Kamiokande have sensitivity beyond that of direct detection experiments. 
The next generation of experiments, that is JUNO, DUNE, and Hyper-Kamiokande, 
should further expand the sensitivity to MDM. A search for multi-hit events 
would probe parameter space for QCD density MDM such as quark nuggets, while a 
background free single event analysis could be sensitive to electroweak density 
MDM such as \EDBs/.

%%%%%%%%%%%%%%%%%%%%%%%%%%%%%%%%%%%%%
% Acknowledgements
%%%%%%%%%%%%%%%%%%%%%%%%%%%%%%%%%%%%%
\subsubsection*{Acknowledgements}
We thank Tingjun Yang for useful discussion.  The work of Y.B. is supported by the U. 
S. Department of Energy under the contract no. DE-SC-0017647. The work of J.B. 
is supported by PITT PACC.

%%%%%%%%%%%%%%%%%%%%%%%%%%%%%%%%%%%%%
% References
%%%%%%%%%%%%%%%%%%%%%%%%%%%%%%%%%%%%%
\bibliographystyle{JHEP}
\bibliography{mdm}

\end{document}